\input{aipcheck}

\documentclass[
    ,final            
  ]
  {aipproc}

\layoutstyle{6x9}

\begin{document}

\title{A study of spectra of Cyg X-3 observed by BeppoSAX}


\author{A. Szostek}{
  address={Centrum Astronomiczne im.\ M. Kopernika, Bartycka 18, Warszawa, Poland}
}
\author{A. A. Zdziarski}{
  address={Centrum Astronomiczne im.\ M. Kopernika, Bartycka 18, Warszawa, Poland}
}

\begin{abstract}
We model the $\sim$1--200 keV spectra of Cygnus X-3 observed by {\it BeppoSAX}. The continuum, modeled by Comptonization in a hybrid plasma, is modified by the strongly ionized plasma of the stellar wind of the Wolf-Rayet companion star. Discrete absorption and emission spectral features are modeled with {\tt XSTAR}. The model has been applied to phase-resolved spectra in the hard and soft spectral states.
\end{abstract}

\maketitle


\section{Introduction}

Cygnus X-3, a high-mass X-ray binary, contains a compact object (either a black hole or a neutron star) and a Wolf-Rayet star. The orbital period is very short, 4.8~h. A strong, dense, stellar wind from the companion strongly modifies the intrinsic continuum by absorption and emission.

High-resolution spectroscopy provides valuable diagnostic tools to examine the
discrete absorption and emission features of the wind. However, the relatively narrow energy band of {\it Chandra\/} does not allow us to model the intrinsic broad-band continuum. On the other hand, the low spectral resolution and the lack of soft X-ray coverage of {\it RXTE\/} do not allow us to study the discrete emission and absorption features imprinted by the wind. Thus, here we take advantage of the data from {\it BeppoSAX}, featuring broad-band coverage and medium energy resolution.

\section{Observations and model}

We analyze 9 out of 10 observations of Cyg X-3 performed by {\it BeppoSAX}, using the MECS2 (1.7--10.5 keV) and PDS (15--120 keV) data. Each observation contained more than four orbits. We use the parabolic ephemeris \cite{Singh} to divide the data into intervals of 1/6 of the period, and then add together the spectra from a given observation corresponding to the same phase interval. The minima and maxima correspond to the phases of $-0.08$--0.08 and 0.42--0.58, respectively.

\begin{figure}[b]
  \resizebox{11pc}{!}{\includegraphics{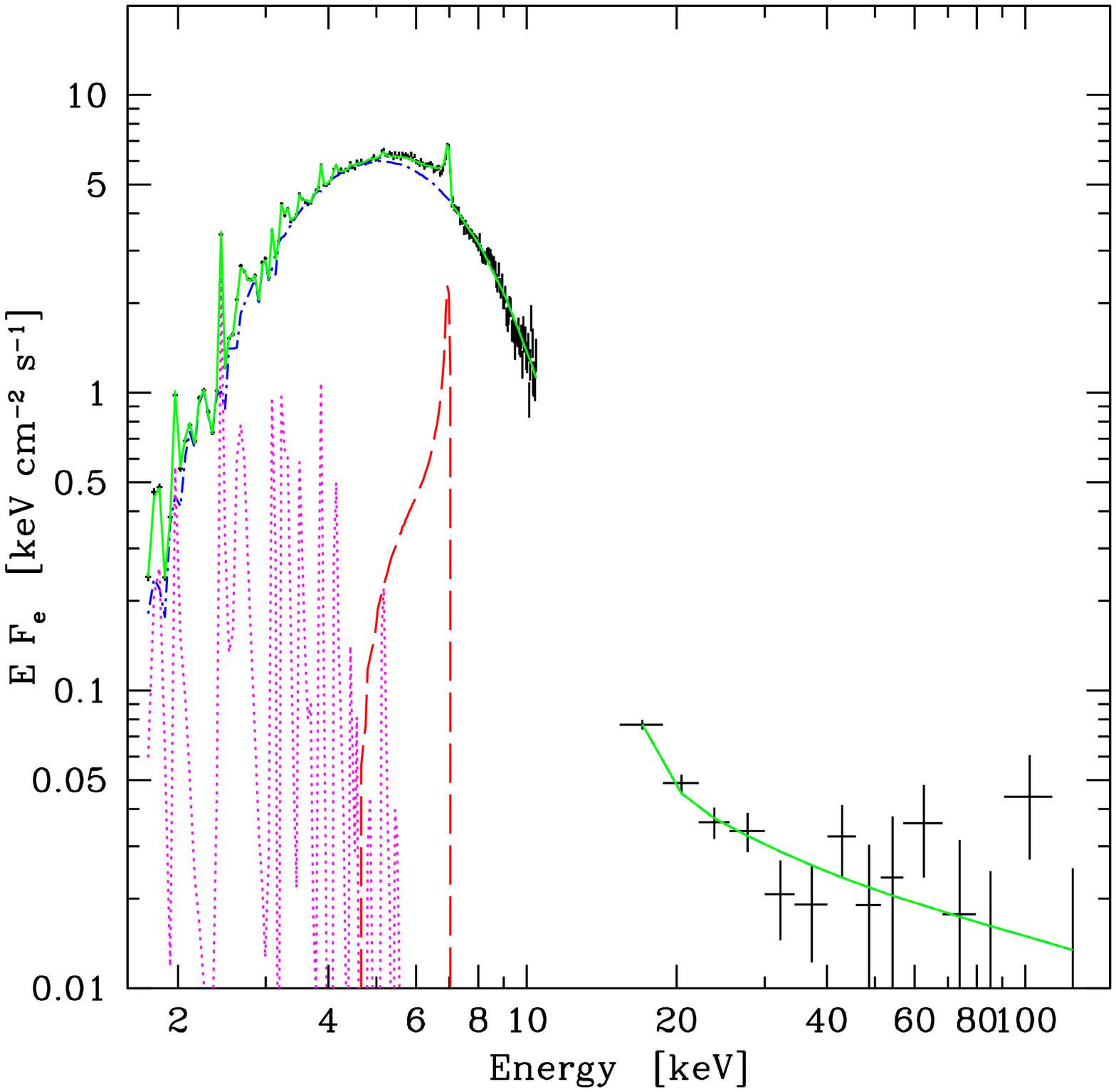}}
  \resizebox{11pc}{!}{\includegraphics{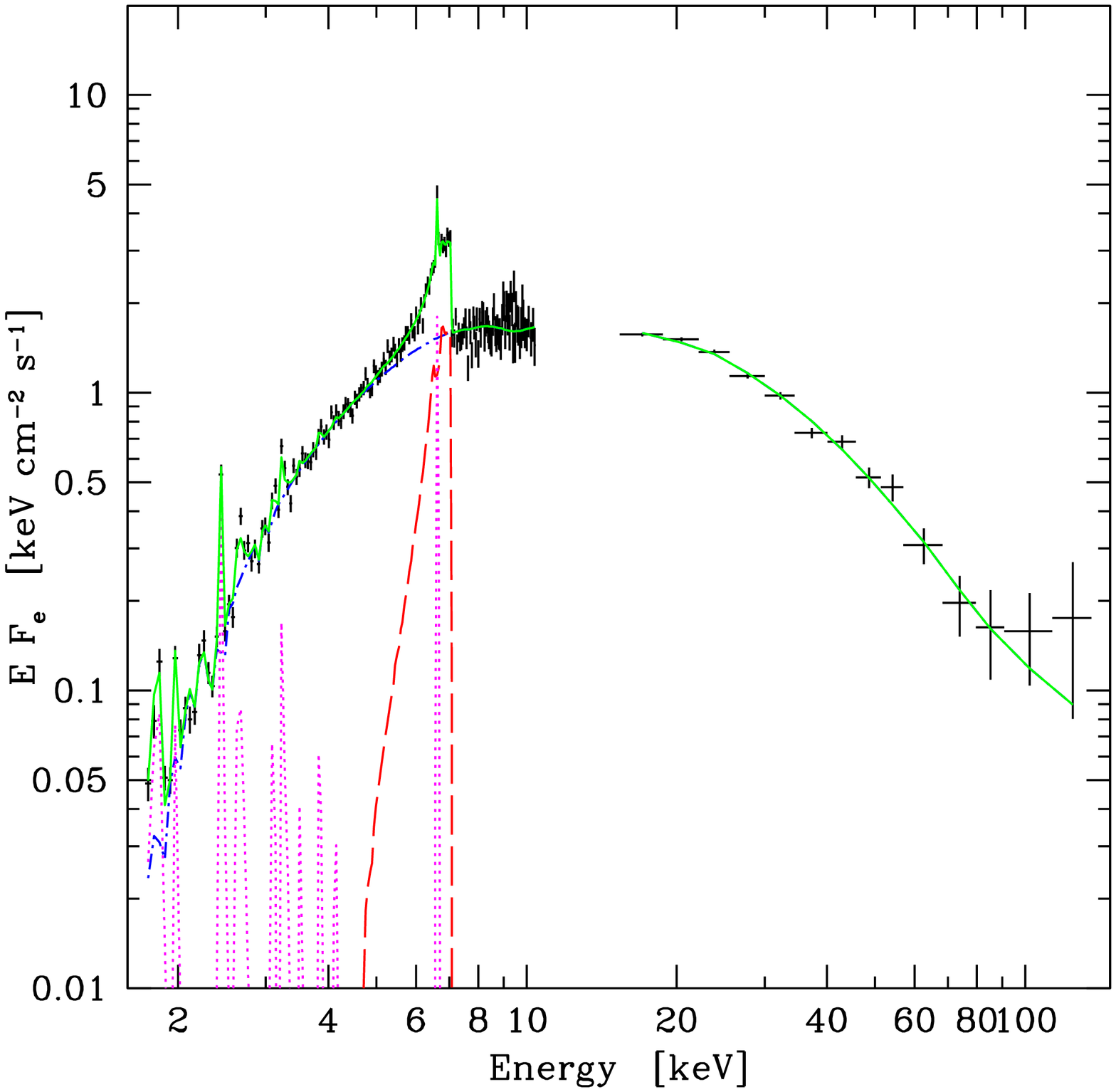}}
\caption{Results of spectral fits. ({\it a}) The softest and ({\it b}) the
hardest observed spectra at the maximum of the orbital modulation plotted
together with the model ({\it solid curves}, {\it green}) consisting of
Comptonization continuum ({\it dot-dashed curves}, {\it blue}) soft X-ray line
emission ({\it dotted curves}, {\it magenta}) and the Fe K line ({\it dashed
curves}, {\it red}).}
\label{Fig1}
\end{figure}

\begin{figure}
 \resizebox{0.31\columnwidth}{!}{\includegraphics{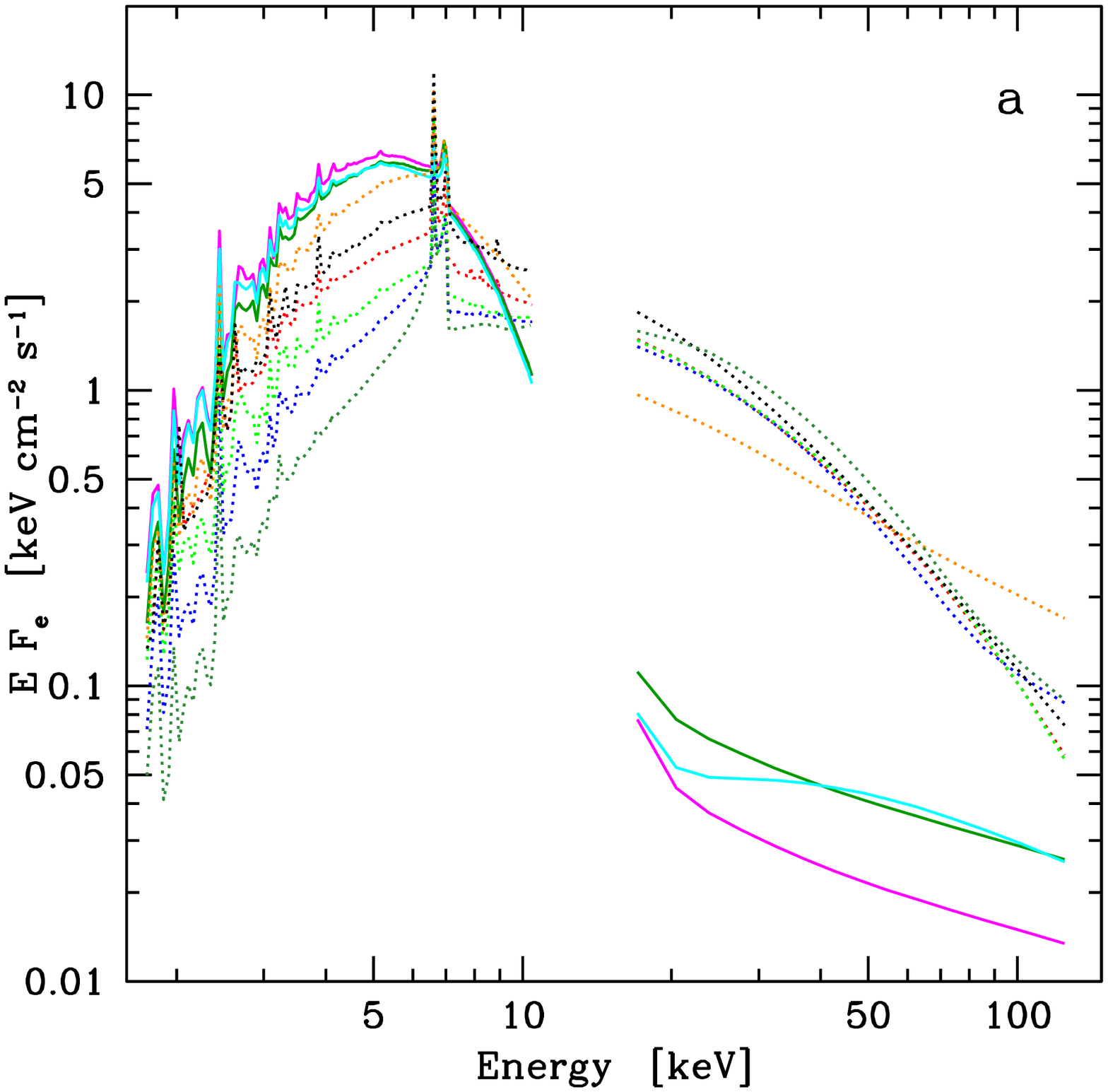}}
 \resizebox{0.31\columnwidth}{!}{\includegraphics{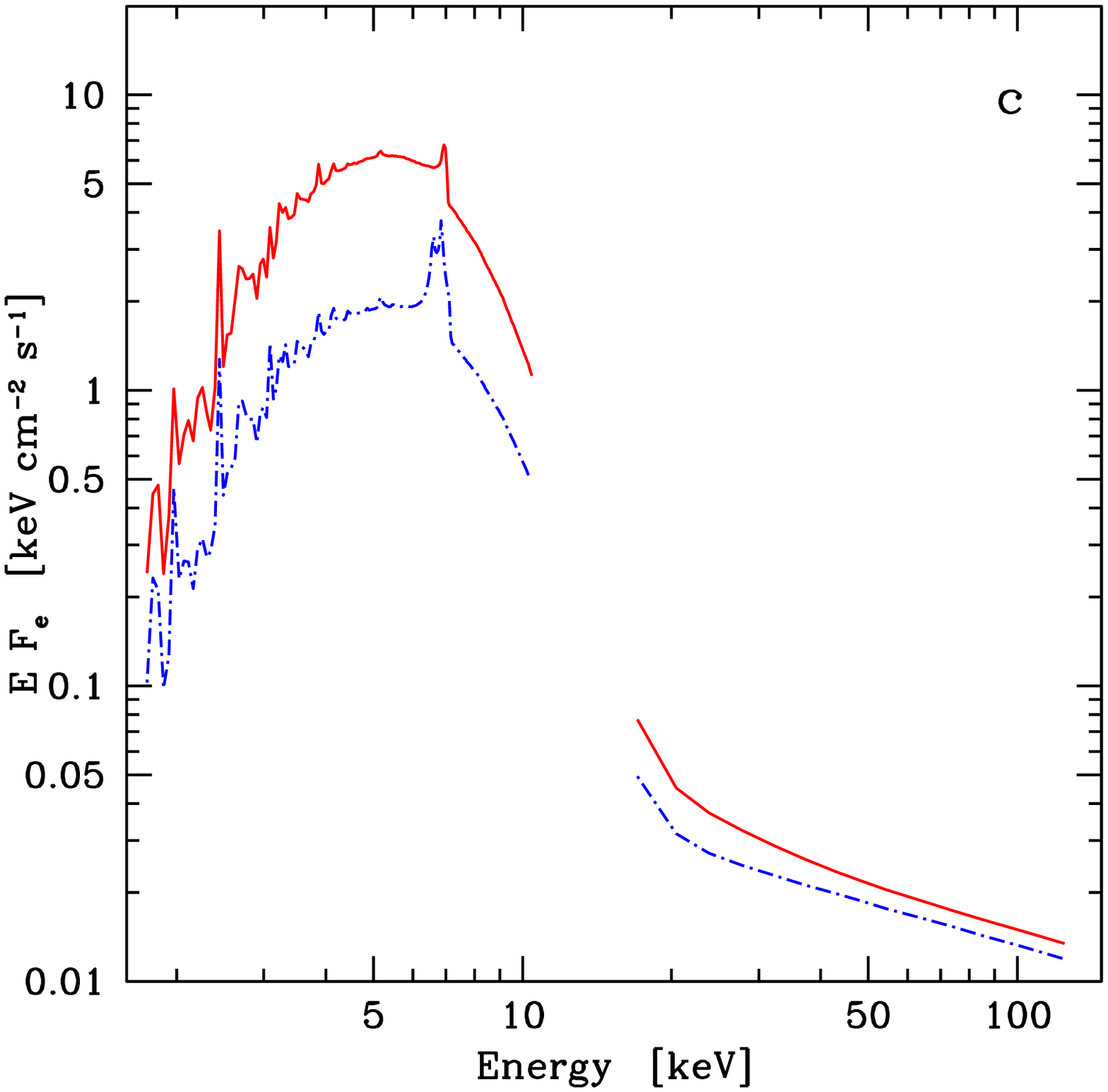}}
\label{Fig2}
\end{figure}
\begin{figure}
\resizebox{0.31\columnwidth}{!}{\includegraphics{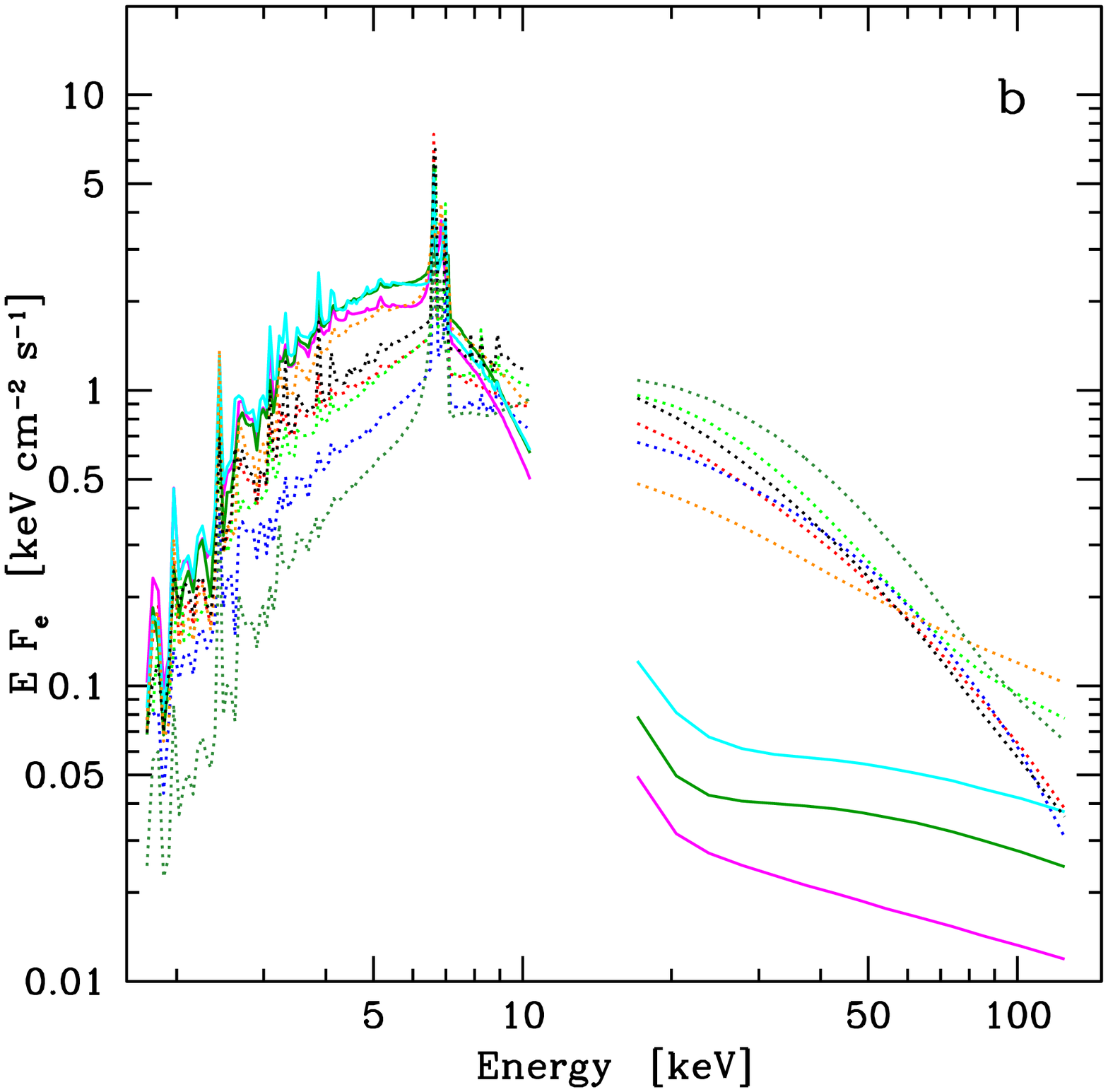}}
 \resizebox{0.31\columnwidth}{!}{\includegraphics{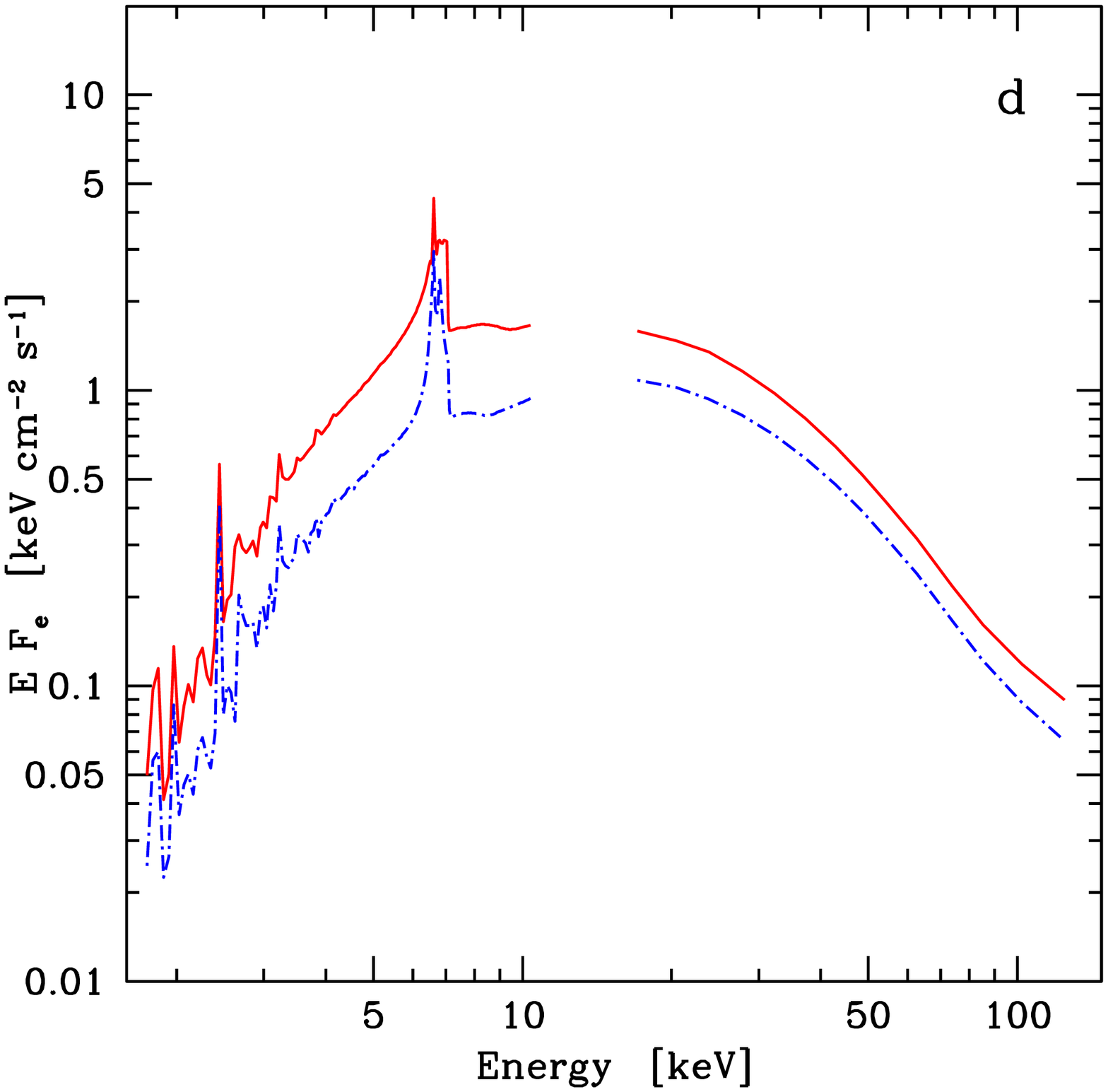}}
\caption{The models fitted to the phase ({\it a}) maxima and ({\it b}) minima
in the soft ({\it solid curves}) and hard ({\it dotted curves}) states, and the
models fitted to ({\it c}) the softest and ({\it d}) the hardest spectra in the
phase minimum ({\it dot-dashed curves}, {\it blue}) and the maximum ({\it solid
curves}, {\it red}).}
\label{Fig2}
\end{figure}

All the spectra have been fitted with a physically-motivated model including
Comptonization of blackbody emission by hybrid (thermal and nonthermal) electrons \cite{Coppi}, \cite{Gierlinski}, Compton reflection from an ionized medium \cite{Magdziarz}, emission and absorption in the photoionized wind calculated with the {\tt XSTAR} photoionization code \cite{Kallman}, neutral absorption, and a relativistically broadened Fe K$\alpha$ line \cite{Fabian}.

\section{Results and conclusions}

The softest and the hardest spectra at the maximum of the orbital modulation 
are plotted in Fig.\ 1. The soft-state spectrum is dominated by a soft blackbody-like component and a high-energy tail due to emission of nonthermal electrons. The hardest spectrum observed by {\it BeppoSAX\/} is well modeled by thermal Comptonization, with no nonthermal electrons required by the fit, and weak Compton reflection. This spectrum is still softer than the hardest spectra observed by {\it RXTE}, similar to the average spectrum 2 in the classification of \cite{Szostek}. Both soft and hard state spectra are significantly absorbed and contain numerous discrete spectral features.

The models fitted to all analyzed spectra in the maxima and minima are shown in Fig.\ 2({\it a}, {\it b}). We clearly see that whereas the spectra below $\sim$10 keV show a continuity of flux levels, the spectra above $\sim$20 keV show a bimodal distribution. Thus, we base our state classification here on the latter, with the soft and hard state corresponding to the low and high, respectively, level of the high-energy flux. (This classification may differ from some other ones used for X-ray binaries.) Interestingly, all the (soft-state) spectra with the weak tail correspond to an almost unique spectrum below 10 keV, forming a thick line in Fig.\ 2(a). The state transition (i.e., the decline of the tail amplitude) takes place only when the soft X-ray flux reaches its very maximum. The soft and hard state spectra intersect at the pivot energy of $\sim$10 keV. 

The amplitude of the orbital modulation, see Fig.\ 2({\it c}, {\it d}), below $\sim$10 keV is comparable with that of the aperiodic variability, see Fig.\ 2({\it a}, {\it b}). On the other hand, X-rays above $\sim$ 20 keV change much less. We find that the flux of the Fe K$\alpha$ line does not vary significantly during either the orbital modulation or state transition. This is suggestive of the origin of this emission from an extended region. Naturally, the equivalent width is the highest at the minimum of the continuum flux, which corresponds to the phase minimum of the hard state. 

Although the energy resolution of {\it BeppoSAX\/} does not allow for precise separation of adjacent spectral features in soft X-rays or precise measurement of their equivalent width, we still have been able to identify several lines and recombination continua such as: Si~{\sc xii}, Si~{\sc xiv}~Ly~$\alpha$, RRC~Si~{\sc xiii}, S~{\sc xvi}~Ly~$\alpha$, Si~{\sc xiv},
Ar~{\sc xvii}, Ar~{xviii}~Ly~$\alpha$, RRC~S~{\sc xvi}, Ca~{\sc xix}.

\begin{theacknowledgments}
We thank L. Hjalmarsdotter for valuable comments on this paper. This research has been supported by KBN grants 1P03D01827, 4T12E04727 and PBZ-KBN-054/P03/2001. 
\end{theacknowledgments}

\end{document}